\documentclass[journal]{IEEEtran}

\usepackage{xcolor,soul,framed}
\colorlet{shadecolor}{yellow}
\usepackage[pdftex]{graphicx}
\graphicspath{{../pdf/}{../jpeg/}}
\DeclareGraphicsExtensions{.pdf,.jpeg,.png}
\usepackage{bbding}

\usepackage[cmex10]{amsmath}

\usepackage{array}
\usepackage{mdwmath}
\usepackage{mdwtab}
\usepackage{eqparbox}
\usepackage{url}
\usepackage{hyperref}
\usepackage{algorithm, algorithmic}
\usepackage{amssymb, amsfonts}
\usepackage{makecell}

\usepackage{threeparttable} 
\usepackage{multirow}
\usepackage{multicol}
\usepackage{booktabs}
\usepackage{subfigure}
\usepackage{hyperref}
\begin{document}
\bstctlcite{IEEEexample:BSTcontrol}
    \title{sEMG-based Gesture-Free Hand Intention Recognition: System, Dataset, Toolbox, and Benchmark Results}
  	\author{Hongxin  Li,
                  Jingsheng Tang*,  
                  Xuechao Xu, 
                  Wei Dai,
                  Yaru Liu,
                  Junhao Xiao,\\
                  Huimin Lu, IEEE Member, 
                  and~Zongtan Zhou
\thanks{This work was supported in part by the Science and Technology Innovation 2030 “Brain Science and Brain-Like Research” Key Project of China under Grant 2022ZD0208504; in part by the National Natural Science Foundation of China under Grant U22A2059, Grant U1913202, and Grant 62203460; and in part by the Major Project of the Natural Science Foundation of Hunan, China, under Grant 2021JC0004.}
\thanks{All authors are with the College of Intelligence Science and Technology, National University of Defense Technology, Changsha 410073, China.}
\thanks{* Corresponding author (e-mail: mrtang@nudt.edu.cn).}
}

% The paper headers
\markboth{Submitted to IEEE Transactions on Industrial Informatics}{Li \MakeLowercase{\textit{Li et al.}}: sEMG-based Gesture-Free Hand Intention Recognition}

% It enables quick and effective communication of complex information within the team and enhances information security.

\maketitle

\begin{abstract}
In sensitive scenarios, such as meetings, negotiations, and team sports, messages must be conveyed without detection by non-collaborators. Previous methods, such as encrypting messages, eye contact, and micro-gestures,  had problems with either inaccurate information transmission or leakage of interaction intentions. To this end, a novel gesture-free hand intention recognition scheme was proposed, that adopted surface electromyography(sEMG) and isometric contraction theory to recognize different hand intentions without any gesture. Specifically, this work includes four parts: (1) the experimental system, consisting of the upper computer software, self-conducted myoelectric watch, and sports platform, is built to get sEMG signals and simulate multiple usage scenarios; (2) the paradigm is designed to standard prompt and collect the gesture-free sEMG datasets. Eight-channel signals of ten subjects were recorded twice per subject at about 5-10 days intervals; (3) the toolbox integrates preprocessing methods (data segmentation, filter, normalization, etc.), commonly used sEMG signal decoding methods, and various plotting functions, to facilitate the research of the dataset; (4) the benchmark results of widely used methods are provided. The results involve single-day, cross-day, and cross-subject experiments of 6-class and 12-class gesture-free hand intention when subjects with different sports motions. To help future research, all data, hardware, software, and methods are open-sourced on the following website: \href{https://tammie-li.github.io/}{\textcolor{red}{click here}}. 

\end{abstract}


\IEEEpeerreviewmaketitle

\section{Introduction}
\IEEEPARstart{I}{n} the scenario where collaborators and non-collaborators coexist, how to achieve message transmission between collaborators without being noticed by non-collaborators is of great significance for team cooperation in conference negotiations or team sports. Previous research has focused on the concealment of interactive information, ensuring that it is difficult for non-collaborators to decode. For example, Katriel et al. \cite{EMG_1} utilize end-to-end encryption and decryption technologies to prevent messages from being stolen and exploited. Another typical approach is to develop a specialized code language for collaborators, such as eye contact \cite{EMG_2} and gestures \cite{EMG_3,jung2015wearable}, that are difficult to understand for non-collaborators. Although the above methods prevent non-collaborators decode the meaning of the message, the process of sending the message is exposed, which may easily arouse the perception and vigilance of non-collaborators. In recent years, extensive research on human-computer interaction has satisfied the concealment, but there are shortcomings in other aspects. Typically, brain-computer interaction directly decodes human intentions for transmitting messages without body movements \cite{EMG_4}, but has problems such as dependence on stimulus paradigms, low accuracy, and poor wearing comfort. Gaze-based interaction only produces slight head displacement, but it is limited by the Midis touch problem \cite{EMG_6} in high-eye load application scenarios. Therefore, it is necessary to design a novel interaction method, that can convey messages accurately without obvious body movements or explicit signals. 

The gesture recognition system based on surface electromyography (sEMG) signals has been widely studied and achieved robust performance\cite{jung2015wearable}. As a weak electrobiological signal, it reflects information related to surface muscle and bone activity in the human body. It is a motor unit action potential train sequence generated by the recruited motor units during muscle excitation, which is filtered by muscles, subcutaneous fat, and skin tissue, and superimposed on the electrode on the skin surface to form a signal. The sEMG signals are the neural electrical signals accompanying muscle contractions, which can truly reflect the intention of the human body to make movements, and the relationship between intention and movement is not sufficient or necessary. A typical case is the isometric contraction of muscles \cite{EMG_7}, where the muscles produce excitation and output force, but their length remains unchanged. The muscle isometric contraction refers to the contraction of muscles that maintain a constant length while experiencing changes in tension. In this contracted state, muscle tension can increase to its maximum. However, due to the absence of displacement, physically speaking, muscles do not perform any external work, yet still require a significant amount of energy consumption. The active contraction of muscle components generates significant tension, which causes the elastic components of the muscle to elongate and counteract. Therefore, regardless of whether the limbs produce movements or not, sEMG signals can always accurately determine whether a person has the intention to control limb movements. This is the theoretical basis for selecting sEMG signals to recognize gesture-free hand intentions in this work. 

Based on the isometric contraction theory, the experimental system is built to acquire sEMG data without any gesture. The system first designs a myoelectric wristband, that can acquire eight-channel sEMG signals and three-axis acceleration signals with a 500Hz sampling rate. The supporting host computer software has two pages, one is to help users control, store, and visualize the data, and another is used to control the experimental paradigm and prompt subjects to apply the correct hand force. Considering users may use this interaction technology in various sports scenarios, we have specially customized a sports platform that supports simulating usage at different sports speeds. 

To standardize the recording process of sEMG, the corresponding experimental paradigm has been designed. The experimental procedure consists of 12 blocks $\times$ 12 trials of data recording. The trail number corresponds one-to-one with the hand intention. Block denotes the current sports state of the subject, which includes three rounds of switching between four different sports states. Each trial records the sEMG signals for ten seconds with the prompts of voice, text, image, and progress bar. Ten subjects take part in the above procedure twice, with an interval of 5-10 days. The data format has 15 channels, which denotes the real-time sEMG data, IMU data, and information on the block, speed, time stamp, and trigger, respectively. 

% for each experiment is $T \times C$, where $T$ is the dot of the experiment time and the sampling rate. $C=15$ records the real-time sEMG data, IMU data, and information on the block, speed, time stamp, and trigger. 

To promote future research, we have developed a toolbox for this dataset that supports the function for data preprocessing, sEMG classification, and visualization. The data preprocessing section converts the collected sEMG dataset into data samples for model training through steps such as data slicing, filtering, baseline correction, and dataset preparation. We have summarized and reproduced the representively sEMG signal decoding methods in recent years, and will use them as the baseline to test the classification performance of the gesture-free hand intention recognition task. Finally, the toolbox provides most of the visualization analyse methods, such as t-distributed stochastic neighbor embedding (t-SNE), confusion matrix, spectrogram, and etc. 

Based on the dataset recorded by the experimental system and paradigm, and the analysis tools and methods supported by the toolbox, the mainstream sEMG classification methods have been tested and the benchmark results for the gesture-free hand intention recognition task have been provided. The experiments involve three types: single-day, cross-day, and cross-subject. To adapt to the different application scenarios, the length of the time window is set to 250ms, 500ms, and 750ms, respectively, The classification categories are set to 6 and 12. The former only includes samples of single-finger force and resting state, while the latter expands samples of multiple-finger combination force. In addition, we further obtained the details of the classification results by the confusion matrix tool and analyzed the impact of different sports modes on the task. Finally, we conducted online experiments to demonstrate the effectiveness of the proposed scheme.

This work first proposes the scheme of gesture-free hand intention recognition based on the sEMG signal and the isometric contraction theory. The detailed contribution and significance of this work lie in:
\begin{itemize}
    \item[1)] Designed a sEMG-based gesture-free hand intention recognition experimental system, which enhances the concealment of information transmission while ensuring accurate and efficient information transmission;
    \item[2)] Based on this system, we designed the paradigm and recorded the gesture-free sEMG dataset containing ten subjects. Considering real-world application scenarios, the data is collected when the subjects are standing, slow, fast, and jogging;
    \item[3)] For the convenience of data analysis, we have developed a data analysis toolbox, which includes functions such as dataset management, data processing, and visualization analysis;
    \item[4)] To promote further research, benchmark results of representative sEMG signal processing algorithms were provided on this dataset. 
    
\end{itemize}

\begin{figure*}[htb]
    \centering
    \centerline{\includegraphics[width=2\columnwidth]{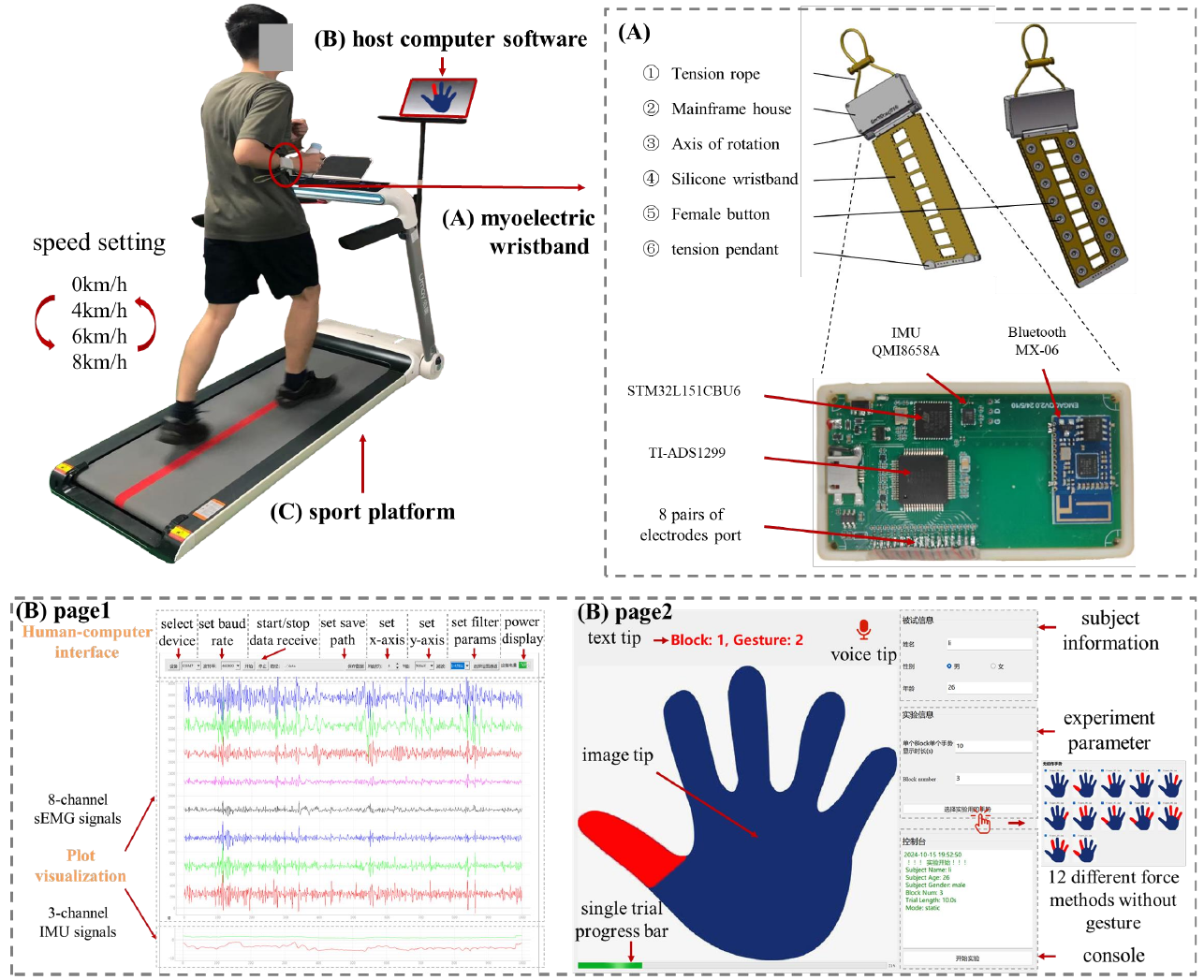}}
    \caption{Experimental system for sEMG-based gesture-free hand intention recognition. It comprises the self-conducted myoelectric wristband, the matched host computer software, and the Umay U3H sports platform. The myoelectric wristband is responsible for recording real-time sEMG signals; The host computer software has two pages, one is designed for data parameter setting and visualization, and another is for paradigm parameter settings and task prompts; The sports platform simulates different application scenarios. }
    \label{fig-1}
\end{figure*}

\section{Experimental System}
As illustrated in Fig. \ref{fig-1}, the experimental system consists of three parts: 1) the self-conducted myoelectric wristband, which is designed for recording the real-time sEMG signals. The sub-figure on the right shows the details of external structure and PCB circuit design; 2) the matched host computer software, consists of two graphical user interface (GUI) pages. The first page is responsible for managing data, including the functions of data receiving, storage, and visualization, while the second page is responsible for managing paradigms, including the functions of paradigm control and task prompts; 3) Umay U3H sports platform, considering the different application scenario, we adopt a treadmill to simulate the various motion patterns possible for the user. The detailed introduction of the first two parts is as follows:

\subsection{Self-conducted myoelectric wristband}
Before designing the myoelectric wristband, we considered the problem of the sEMG source, which part of the body is used to collect the signals. Due to the well-developed muscle group in the forearm, previous studies have often recorded sEMG signals in this area, but this may not be effective for gestures involving fine finger movements. Botros et al. \cite{EMG_8, jiang2017feasibility} show that classifiers trained and tested using wrist EMG signals are significantly higher than the former in both single-finger and multi-finger gestures. This is because controlling the muscle group extension range of human fingers covers the wrist, which means that electromyographic signals representing human finger movement intention can be effectively collected in the wrist, demonstrating the theoretical basis for realizing gesture-free hand intention recognition. In addition, using wrist EMG signals can be more easily integrated into wristband devices, making consumers wear more comfortable. 

From part A of Fig. \ref{fig-1}, the myoelectric wristband consists of two parts: the host and the strap. The host is composed of a PCB circuit board and a corresponding outer shell, while the strap is made of soft silicone. The wires between each electrode and the host are hidden in an inner silicone composite molding technology cavity. The spindle adopts a hollow design, and the cables in the silicone band are connected to the PCB main board of the main body casing through the hollow spindle. Eight pairs of 4.0mm medical standard female buckles are arranged longitudinally on the inside of the wristband, which is compatible to be used with electrode pads of 3.9mm standard interface. Considering the requirement for low power consumption, the STM32L151CBU6 chip is selected to control the data acquisition and forwarding process. In terms of analog-to-digital converters (ADC), we have chosen ADS1299 from Texas Instruments (TI) company to collect weak surface electromyography signals, which is an integrated acquisition chip specifically designed for physiological electrical signals. Its features include an eight-channel 24 bit ADC analog-to-digital sampling module with a programmable gain programmable gain amplifier (PGA) added to the front-end, as well as a right leg drive circuit for denoising. In addition, a three-axis accelerometer supported by QMI8658A is built to measure x, y, and z three-dimensional acceleration signals. Based on the size of the comprehensive data transmission, speed requirements, and power consumption, the system selects MX-06 and uses Bluetooth SPP protocol to transmit the sEMG data to the host computer software. Overall, the self-conducted myoelectric wristband supported to collect and transmit the eight-channel wrist sEMG signals for monitoring the tension/relaxation state of the hand muscles or the intention of the movement, and 2) three-axis acceleration signals for monitoring the movement state of the wearer. 

\subsection{Design of the host computer software}
From part B of Fig. \ref{fig-1}, the host computer consists of two pages that switch with a button: (1) data manager; and (2) paradigm manager, which uses a button to switch. These graphical user interfaces are developed to receive the sEMG data,  visualize the multi-channel signals, and record necessary experimental information. The following is a detailed introduction to the functions of each page:

\subsubsection{Page1: data manager}
\begin{figure}
    \centering
    \centerline{\includegraphics[width=\columnwidth]{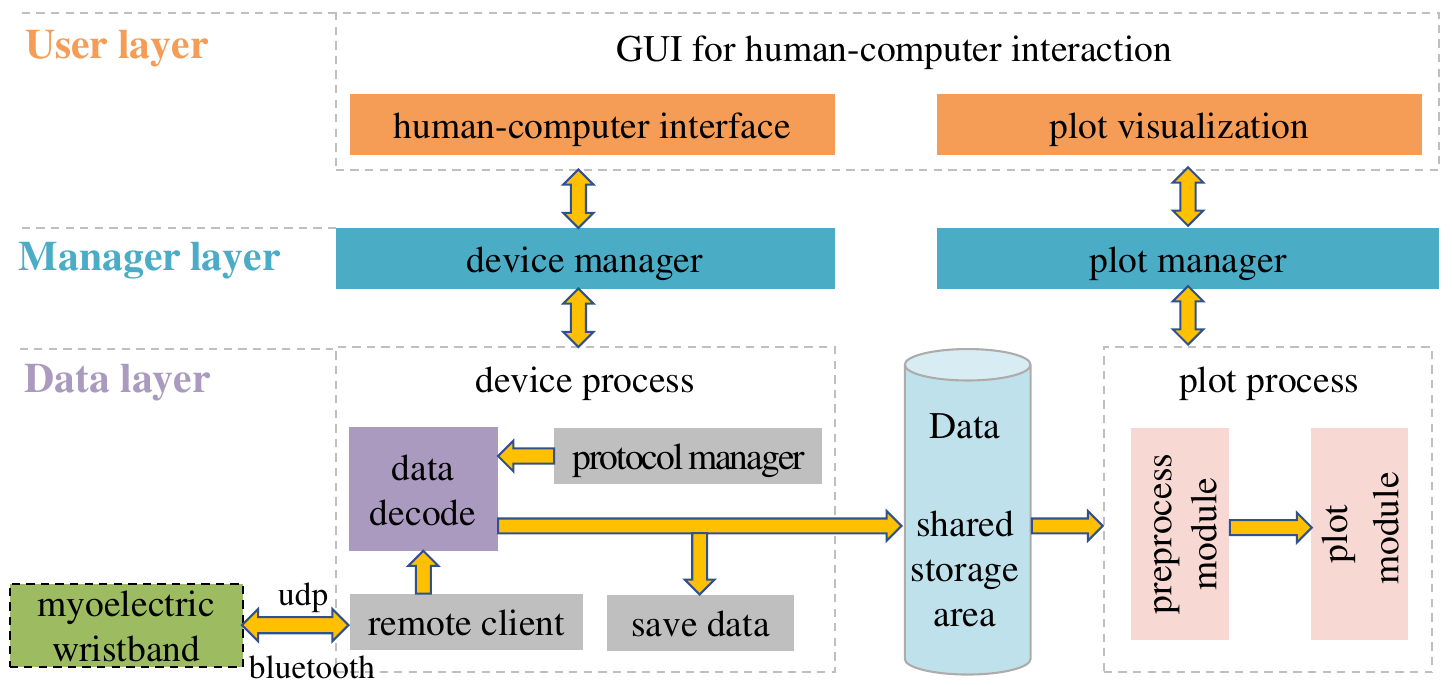}}
    \caption{Software architecture of the page1: data manager, which consists of three layers, i.e. the user layer for human-computer interface; the manager layer controls device and plot operations; and the data layer for data save, decode, preprocess, and plot.}
    \label{fig-host_ar}
\end{figure}

The host computer has three real-time functions, i.e. receiving the byte stream data from the myoelectric wristband; decoding the sEMG signals based on the data protocol; and visualizing the multi-channel signals. Fig. \ref{fig-host_ar} shows the hierarchical software architecture of this page, which consists of user, manager, and data layers. The GUI of the user layer is shown in Fig. \ref{fig-1}(B1), implemented based on the PyQt library. The top column is operable and used to set parameters for various functions. The following two columns cannot be operated, they are used to display the waveforms of sEMG signals and acceleration signals, respectively. All functions are divided into device functions and plot functions, running on two different threads and sharing data through memory-sharing technology. To facilitate the management of functions, we have set up two managers to implement the classification management and calling of functions.

\subsubsection{Page2: paradigm manager}
This page is designed for paradigm parameter settings and task prompts. To facilitate information management and assist subjects in accurately completing experimental tasks, we have developed a graphical interface that prompts subjects to switch and maintain the corresponding gesture-free hand intention state from four aspects: images, text, sound, and progress bar. The GUI is shown in Fig. \ref{fig-1}(B2), with the experimental prompt area on the left and the experimental settings area on the right.

\section{Paradigm and Dataset}
Based on the above experimental system, an experimental paradigm was designed to construct the sEMG-based gesture-free hand intention recognition dataset. The paradigm collected eight-channel sEMG data and three-channel accelerometer signals with a 500Hz sampling rate from ten subjects.

\subsection{Subjects}
Before each experiment procedure, the subject needs to fill in personal information in the upper right corner of Fig. \ref{fig-1}(B2). According to the statistics, ten subjects (6 males and 4 females, aged between 22-28 years old) participated in our experiments. 
% All subjects were students from the National University of Defense Technology and had no neurological disorders.
All subjects had no muscle or nervous system injuries. 
The experiment strictly adheres to the provisions of the Helsinki Declaration, providing a detailed introduction to the experimental process and signing an informed consent form before the experiment begins. It should be noted that each subject is required to complete two experiments, with an interval of 5-10 days, to construct a cross-day dataset. 

\begin{figure}[htb]
    \centering
    \centerline{\includegraphics[width=\columnwidth]{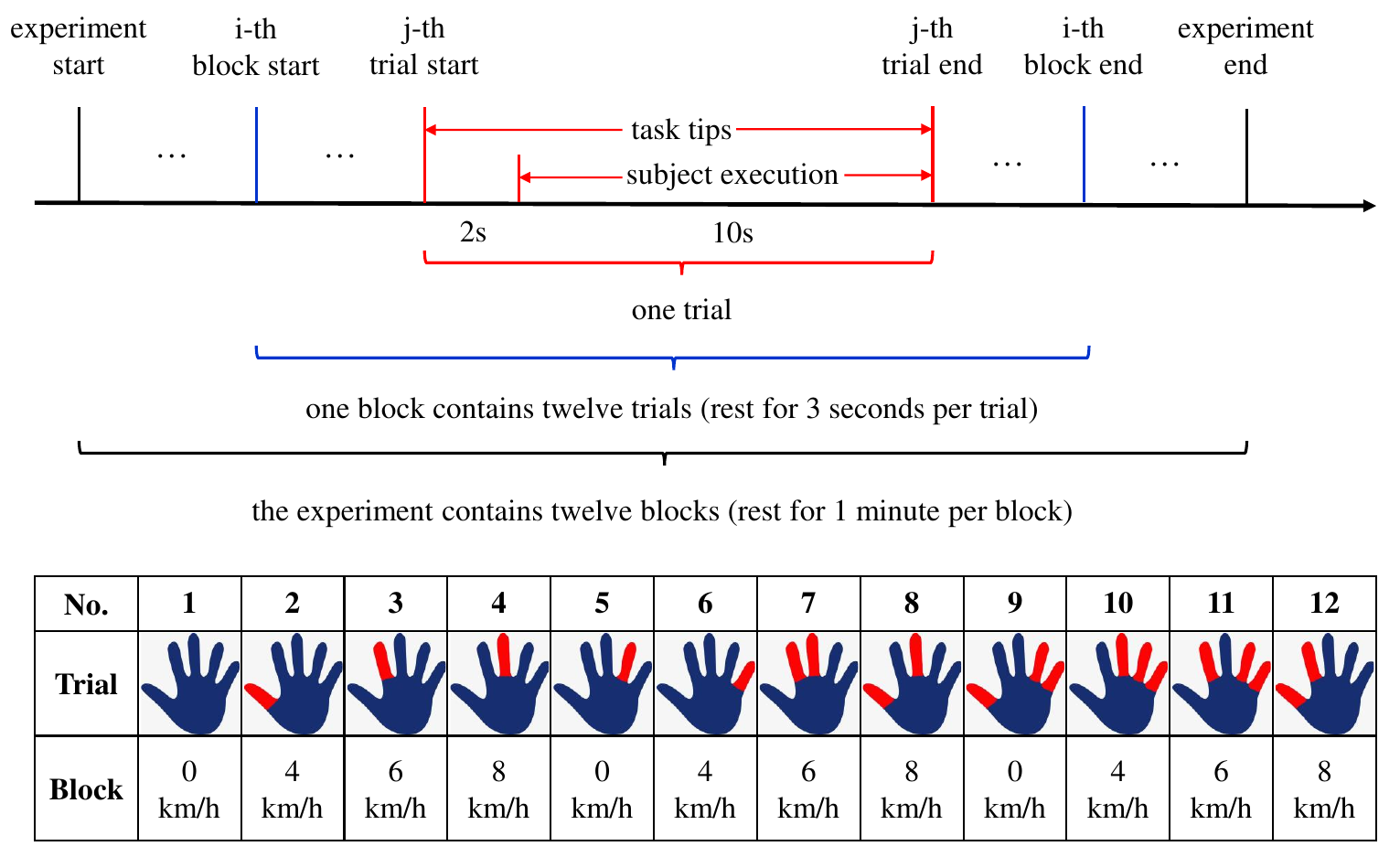}}
    \caption{The sEMG-based gesture-free hand intention recognition paradigm. Each experiment includes 12 blocks and each block includes 12 trials. The motion speed corresponds to each block and the hand force mode corresponds to each trial are different. The specific correspondence is shown in the table.}
    \label{fig-procedure}
\end{figure}

\subsection{Procedure}
Before the paradigm, some preparatory work needs to be completed, such as checking the experimental equipment, setting the host computer software parameters correctly, cleaning the subjects' hands, and wearing the self-conducted myoelectric wristband. Like a smartwatch, the myoelectric wristband is placed directly above the back of the wrist. The experiment consists of twelve blocks, and each block contains twelve trials. In this work, we designed twelve types of gesture-free hand intention states, each corresponding to a trial. Considering the application requirements of decoding intention without gesture in sports scenes, the paradigm sets motion speeds of 0km/h, 4km/h, 6km/h, and 8km/h to simulate four motion states: stationary, slow walking, fast walking, and jogging. The experimental system would select corresponding image tips and motion platform speed settings based on the block and trial number, as shown in the table below Fig. \ref{fig-procedure}. The fingers marked in red indicate that this gesture requires the finger to maintain a vigorous state, while the fingers marked in blue remain relaxed. During the experiment, the subjects were required to hold a cylindrical object (simulate holding things in hand) with their hands and maintain a tight grip on the object with their fingers during the force application process to ensure that sEMG signals were collected gesture-free.

\subsection{Dataset format}
\begin{figure}[htb]
    \centering
    \centerline{\includegraphics[width=\columnwidth]{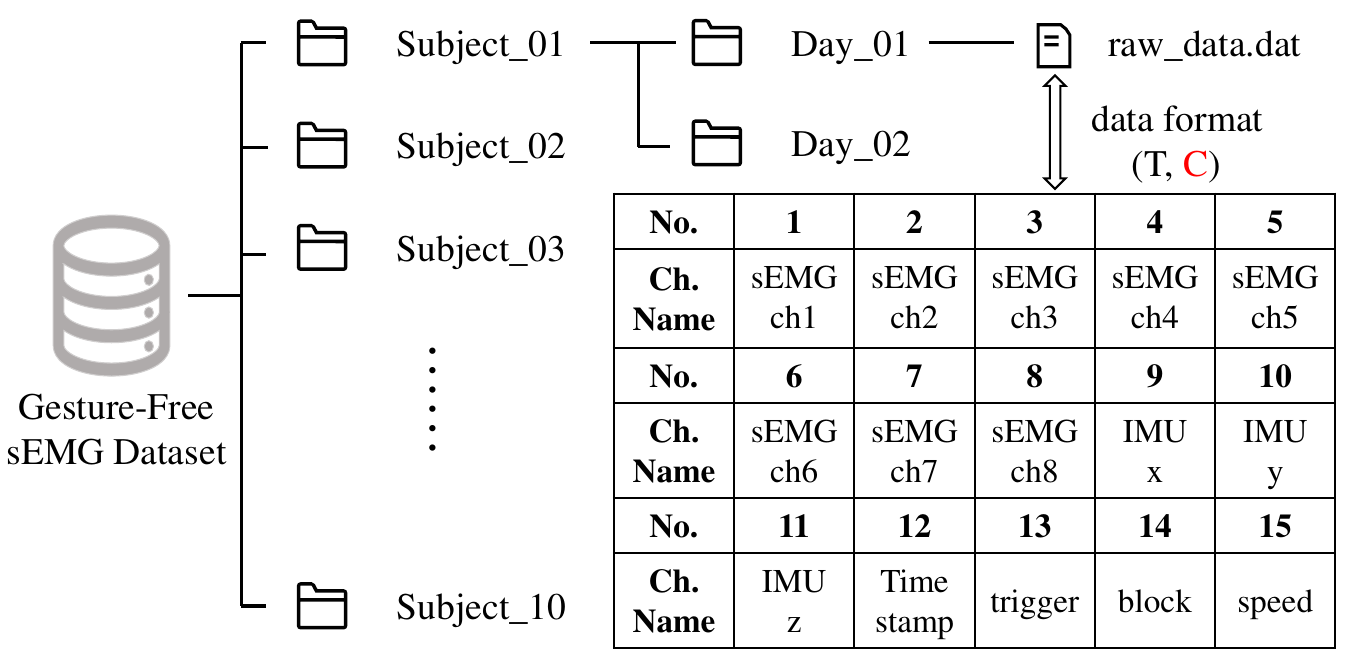}}
    \caption{Dataset format. The dataset includes data from two experiments involving ten subjects. Each experiment records a source file of shape ($T$, $C$), where $T$ and $C$ denote the number of sampling points and channels respectively. There are a total of 15 channels, and the meaning of each channel is shown in the table.}
    \label{fig-format}
\end{figure}

The sEMG-based gesture-free hand intention recognition dataset includes two experiments from ten subjects in two days. In each experiment, the system records a source file of shape ($T$, $C$) with a .dat suffix. The parameter $T$ denotes the sampling points during the experiments, which is the product of recording time and sampling rate. At each sampling point, the experimental system would record 15-channel signals. As shown in Fig. \ref{fig-format}, channels 1-8 record the eight-channel sEMG signals, and channels 9-11 record the axis-x, axis-y, and axis-z of the IMU data. Channel 12 is the sampling timestamp of the data for these 15 channels. Channel 13-15 denote the parameters of the paradigm, i.e. trigger, block, and speed information.

\section{Data Analyse Toolbox}
\subsection{Preprocess Tool}
 To help future researchers in processing and analyzing data, we have developed multiple data preprocessing methods, which are described in detail as follows:
 \begin{itemize}
     \item [1)] \textbf{Data segmentation}: Obtain all sample data that meet the conditions from the raw data based on the label of the gesture, as well as the set time window size and window movement step size. The function head code can be expressed as "def data$\_$segmentation(raw$\_$data, gesture$\_$id, window$\_$size, step)"; 
     \item [2)] \textbf{Data filter}: According to the cutoff frequency, various common filters such as low-pass, high pass, band-pass, and notch are provided. It should be noted that all filters are zero phases to ensure the integrity of phase information. The pseudo-code of the function head is "def data$\_$\{filtername\}$\_$filter(raw$\_$data, [cut$\_$off$\_$params])";
     \item [3)] \textbf{Baseline correction}: sEMG data is non-stationary and exhibits significant time drift. To solve this problem, baseline correction methods are often used to ensure distribution consistency between signals collected at different times. The calculation method is to subtract the signal mean in a calm state at the beginning of the trial from the current time data. The pseudo-code of the function head is "def data$\_$baseline$\_$correction(raw$\_$data, baseline$\_$len)";
     \item [4)] \textbf{Train / test dataset preparation}: This function manage all  recorded sEMG data. Based on the type of experiment, such as single-day, cross-day, or cross-subject experiments, it would produce the training and testing datasets, which are required for the decoding algorithm. The pseudo-code of the function head is "def data$\_$divide$\_$\{experiment-type\}(dataset, [subject$\_$id, day$\_$id, and etc.])".
\end{itemize}

\subsection{Decoding Method}
By researching existing literature, we have implemented the following ten widely used sEMG signal decoding methods. The traditional methods are LDA \cite{antuvan2019lda}), NaiveBayes \cite{EMG_10}, k-nearest neighbor (KNN) \cite{EMG_11}, SVM \cite{alkan2012identification}, and Random Forests \cite{gokgoz2015comparison} and the deep learning methods are LSTM \cite{samadani2018gated}, 1DCNN \cite{cheng2021gesture}, 2DCNN \cite{du2017surface}, CNN-LSTM \cite{xie2018movement}), and L-EMGNet \cite{EMG_9}, respectively. For traditional classification methods, the time-frequency characteristics of sEMG signals are extracted as classification features, including signal mean, root mean square (RMS) value, average frequency, and spectral density. It should be noted that all methods can run end-to-end to meet practical application needs, and the code is open-sourced on the project website.

\subsection{Visualization Tool}
Data visualization can help us assess signal quality, discover objective patterns, and further optimize research. For this purpose, we have developed various function scripts and implemented commonly used visualization methods, including (1) 1D / 2D waveform plotting, for observing and comparing sEMG temporal dynamics; (2) The spectrum diagram used to analyze the main frequency components of sEMG signals and design filters accordingly; (3) confusion matrix is used to analyze the details of algorithm classification results, to screen out the most discriminative force application method; (4) t-SNE distribution, used to observe the distribution of sEMG features, and etc. 

\begin{table*}[htb]
	\centering
	\caption{Benchmark Results of Single-day, Cross-day, and Cross-subject Experiments with Different Window Sizes (Mean ± Standard Deviation). The Basic and Advanced Gesture-Free Hand Intention Recognition Tasks Include 6 and 12 Power Modes Respectively. The Lengths of the Time Windows Are 250ms, 500ms, and 750ms, Respectively. }
        \resizebox{\textwidth}{!}{
		\begin{threeparttable}
                \label{table_single_day}
			\begin{tabular}{cc|ccc|ccc}
				\toprule 
				\multirow{2}{*}{\textbf{Method}}              &\multirow{3}{*}{\textbf{Type}}  &        
                    \multicolumn{3}{c|}{\textbf{6-classes}}     & \multicolumn{3}{c}{\textbf{12-classes}}   \\ \cmidrule{3-8}
			 \textbf{(Year)} &&  \textbf{250ms}         & \textbf{500ms} &  \textbf{750ms}   & \textbf{250ms}         & \textbf{500ms} &  \textbf{750ms}                        \\
			\midrule
			  \multirow{2}{*}{LDA}   & SD                  & 0.8618 $\;\pm\;$0.0807\;    
                                                             & 0.8877 $\;\pm\;$0.0769\;        
                                                             & 0.8925 $\;\pm\;$0.0834\;         
                                                             & 0.6227 $\;\pm\;$0.0921\;          
                                                             & 0.6690 $\;\pm\;$0.0956\;     
                                                             & 0.6869 $\;\pm\;$0.0953\;           
                                                             \\
                                                      & CD   & 0.6352 $\;\pm\;$0.1776\;    
                                                             & \textcolor{blue}{0.6572 $\;\pm\;$0.1815\;}        
                                                             & \textcolor{blue}{0.6590 $\;\pm\;$0.1818\;}         
                                                             & 0.3885 $\;\pm\;$0.1354\;          
                                                             & 0.4066 $\;\pm\;$0.1442\;     
                                                             & \textcolor{blue}{0.4113 $\;\pm\;$0.1459\;}           
                                                             \\
                                                (2019)& CS   & 0.5546 $\;\pm\;$0.1608\;    
                                                             & 0.5700 $\;\pm\;$0.1604\;        
                                                             & 0.5755 $\;\pm\;$0.1623\;         
                                                             & 0.2978 $\;\pm\;$0.0936\;          
                                                             & 0.3100 $\;\pm\;$0.0984\;     
                                                             & 0.3070 $\;\pm\;$0.1000\;           
                                                             \\\midrule
			  \multirow{2}{*}{NaiveBayes}   & SD           & 0.7438 $\;\pm\;$0.1105\;    
                                                             & 0.7657 $\;\pm\;$0.1160\;        
                                                             & 0.7778 $\;\pm\;$0.1178\;         
                                                             & 0.4668 $\;\pm\;$0.1029\;          
                                                             & 0.5055 $\;\pm\;$0.1099\;     
                                                             & 0.5250 $\;\pm\;$0.1114\;           
                                                             \\
                                                      & CD   & 0.5675 $\;\pm\;$0.1769\;    
                                                             & 0.5712 $\;\pm\;$0.1827\;        
                                                             & 0.5709 $\;\pm\;$0.1838\;         
                                                             & 0.3441 $\;\pm\;$0.1224\;          
                                                             & 0.3564 $\;\pm\;$0.1319\;     
                                                             & 0.3618 $\;\pm\;$0.1408\;           
                                                             \\
                                                (2021)& CS   & 0.4878 $\;\pm\;$0.1413\;    
                                                             & 0.5122 $\;\pm\;$0.1498\;        
                                                             & 0.5090 $\;\pm\;$0.1302\;         
                                                             & 0.2566 $\;\pm\;$0.0978\;          
                                                             & 0.2649 $\;\pm\;$0.1022\;     
                                                             & 0.2674 $\;\pm\;$0.1025\;           
                                                             \\\midrule
			  \multirow{2}{*}{KNN}   & SD                  & 0.7459 $\;\pm\;$0.1502\;    
                                                             & 0.7554 $\;\pm\;$0.1542\;        
                                                             & 0.7557 $\;\pm\;$0.1555\;         
                                                             & 0.4838 $\;\pm\;$0.1281\;          
                                                             & 0.5094 $\;\pm\;$0.1401\;     
                                                             & 0.5102 $\;\pm\;$0.1433\;           
                                                             \\
                                                      & CD   & 0.5072 $\;\pm\;$0.1608\;    
                                                             & 0.5040 $\;\pm\;$0.1716\;        
                                                             & 0.5005 $\;\pm\;$0.1721\;         
                                                             & 0.2858 $\;\pm\;$0.1059\;          
                                                             & 0.2908 $\;\pm\;$0.1228\;     
                                                             & 0.2881 $\;\pm\;$0.1234\;           
                                                             \\
                                                (2024)& CS   & 0.4624 $\;\pm\;$0.1158\;    
                                                             & 0.4674 $\;\pm\;$0.1184\;        
                                                             & 0.4612 $\;\pm\;$0.1130\;         
                                                             & 0.2231 $\;\pm\;$0.0535\;          
                                                             & 0.2264 $\;\pm\;$0.0541\;     
                                                             & 0.2251 $\;\pm\;$0.0551\;           
                                                             \\\midrule
			  \multirow{2}{*}{SVM}   & SD                  & 0.7952 $\;\pm\;$0.1215\;    
                                                             & 0.8105 $\;\pm\;$0.1252\;        
                                                             & 0.8114 $\;\pm\;$0.1270\;         
                                                             & 0.5520 $\;\pm\;$0.1291\;          
                                                             & 0.5845 $\;\pm\;$0.1392\;     
                                                             & 0.5899 $\;\pm\;$0.1426\;           
                                                             \\
                                                        & CD & 0.5511 $\;\pm\;$0.1668\;    
                                                             & 0.5495 $\;\pm\;$0.1693\;        
                                                             & 0.5476 $\;\pm\;$0.1762\;         
                                                             & 0.3354 $\;\pm\;$0.1302\;          
                                                             & 0.3359 $\;\pm\;$0.1352\;     
                                                             & 0.3359 $\;\pm\;$0.1376\;           
                                                             \\
                                                (2012)& CS   & 0.5161 $\;\pm\;$0.1181\;    
                                                             & 0.5199 $\;\pm\;$0.1304\;        
                                                             & 0.5304 $\;\pm\;$0.1254\;         
                                                             & 0.2499 $\;\pm\;$0.0751\;          
                                                             & 0.2562 $\;\pm\;$0.0798\;     
                                                             & 0.2575 $\;\pm\;$0.0787\;           
                                                             \\\midrule

			  \multirow{2}{*}{\makecell{Random\\ Forests}}   & SD                   & \textcolor{blue}{0.8872 $\;\pm\;$0.0633\;}    
                                                             & \textcolor{blue}{0.9018 $\;\pm\;$0.0627\;}        
                                                             & \textcolor{blue}{0.9050 $\;\pm\;$0.0666\;}         
                                                             & 0.6577 $\;\pm\;$0.0916\;          
                                                             & 0.6941 $\;\pm\;$0.0959\;     
                                                             & 0.7129 $\;\pm\;$0.0931\;           
                                                             \\
                                                      & CD   & \textcolor{green}{0.6685 $\;\pm\;$0.2080\;}    
                                                             & \textcolor{green}{0.6814 $\;\pm\;$0.2174\;}        
                                                             & \textcolor{green}{0.6861 $\;\pm\;$0.2169\;}         
                                                             & \textcolor{green}{0.4361 $\;\pm\;$0.1639\;}          
                                                             & \textcolor{green}{0.4519 $\;\pm\;$0.1725\;}     
                                                             & \textcolor{red}{0.4697 $\;\pm\;$0.1741\;}           
                                                             \\
                                                (2015)& CS   & \textcolor{blue}{0.5808 $\;\pm\;$0.1290\;}    
                                                             & \textcolor{blue}{0.5933 $\;\pm\;$0.1388\;}        
                                                             & 0.5988 $\;\pm\;$0.1300\;         
                                                             & \textcolor{green}{0.3184 $\;\pm\;$0.0867\;}          
                                                             & 0.3275 $\;\pm\;$0.0897\;     
                                                             & 0.3253 $\;\pm\;$0.0876\;           
                                                             \\\midrule
			  \multirow{2}{*}{LSTM}   & SD                 & 0.8677 $\;\pm\;$0.0709\;    
                                                             & 0.8907 $\;\pm\;$0.0610\;        
                                                             & 0.8824 $\;\pm\;$0.0744\;         
                                                             & \textcolor{blue}{0.6639 $\;\pm\;$0.1003\;}          
                                                             & \textcolor{blue}{0.7216 $\;\pm\;$0.0965\;}     
                                                             & \textcolor{blue}{0.7304 $\;\pm\;$0.1022\;}           
                                                             \\
                                                      & CD   & 0.5963 $\;\pm\;$0.1553\;    
                                                             & 0.6030 $\;\pm\;$0.1587\;        
                                                             & 0.6069 $\;\pm\;$0.1795\;         
                                                             & 0.3862 $\;\pm\;$0.1357\;          
                                                             & 0.3936 $\;\pm\;$0.1557\;     
                                                             & 0.3884 $\;\pm\;$0.1709\;           
                                                             \\
                                                (2018)& CS   & 0.5734 $\;\pm\;$0.1285\;    
                                                             & \textcolor{green}{0.6066 $\;\pm\;$0.1236\;}        
                                                             & \textcolor{red}{0.6192 $\;\pm\;$0.1206\;}         
                                                             & 0.3044 $\;\pm\;$0.0797\;          
                                                             & 0.3287 $\;\pm\;$0.0863\;     
                                                             & \textcolor{blue}{0.3485 $\;\pm\;$0.1024\;}           
                                                             \\\midrule
			  \multirow{2}{*}{1DCNN}   & SD                & 0.7521 $\;\pm\;$0.1168\;    
                                                             & 0.7692 $\;\pm\;$0.1194\;        
                                                             & 0.7722 $\;\pm\;$0.1335\;         
                                                             & 0.4911 $\;\pm\;$0.1163\;          
                                                             & 0.5230 $\;\pm\;$0.1328\;     
                                                             & 0.5205 $\;\pm\;$0.1280\;           
                                                             \\
                                                      & CD   & 0.4991 $\;\pm\;$0.1226\;    
                                                             & 0.5257 $\;\pm\;$0.1493\;        
                                                             & 0.5236 $\;\pm\;$0.1456\;         
                                                             & 0.3155 $\;\pm\;$0.1104\;          
                                                             & 0.3204 $\;\pm\;$0.1341\;     
                                                             & 0.3113 $\;\pm\;$0.1189\;           
                                                             \\
                                                (2017)& CS   & 0.4873 $\;\pm\;$0.1234\;    
                                                             & 0.5087 $\;\pm\;$0.1250\;        
                                                             & 0.4952 $\;\pm\;$0.1151\;         
                                                             & 0.2149 $\;\pm\;$0.0483\;          
                                                             & 0.2127 $\;\pm\;$0.0497\;     
                                                             & 0.2064 $\;\pm\;$0.0515\;           
                                                             \\\midrule
			  \multirow{2}{*}{2DCNN}   & SD                & \textcolor{green}{0.9047 $\;\pm\;$0.0595\;}    
                                                             & \textcolor{red}{0.9273 $\;\pm\;$0.0625\;}        
                                                             & \textcolor{green}{0.9242 $\;\pm\;$0.0830\;}         
                                                             & \textcolor{green}{0.6971 $\;\pm\;$0.1075\;}          
                                                             & \textcolor{red}{0.7439 $\;\pm\;$0.0951\;}     
                                                             & \textcolor{green}{0.7542 $\;\pm\;$0.1098\;}           
                                                             \\
                                                      & CD   & \textcolor{blue}{0.6409 $\;\pm\;$0.1902\;}    
                                                             & 0.6248 $\;\pm\;$0.1970\;        
                                                             & 0.6209 $\;\pm\;$0.2197\;         
                                                             & \textcolor{blue}{0.4002 $\;\pm\;$0.1753\;}          
                                                             & \textcolor{blue}{0.4214 $\;\pm\;$0.1839\;}     
                                                             & 0.4082 $\;\pm\;$0.1658\;           
                                                             \\
                                                (2021)& CS   & \textcolor{red}{0.6044 $\;\pm\;$0.1477\;}    
                                                             & \textcolor{red}{0.6216 $\;\pm\;$0.1575\;}        
                                                             & \textcolor{green}{0.6138 $\;\pm\;$0.1411\;}         
                                                             & \textcolor{red}{0.3222 $\;\pm\;$0.1174\;}          
                                                             & \textcolor{red}{0.3327 $\;\pm\;$0.1268\;}     
                                                             & \textcolor{green}{0.3517 $\;\pm\;$0.1367\;}           
                                                             \\\midrule
			  \multirow{2}{*}{CNN-LSTM}   & SD             & 0.8420 $\;\pm\;$0.1704\;    
                                                             & 0.8507 $\;\pm\;$0.1708\;        
                                                             & 0.8515 $\;\pm\;$0.1730\;         
                                                             & 0.6355 $\;\pm\;$0.1253\;          
                                                             & 0.6771 $\;\pm\;$0.1111\;     
                                                             & 0.6604 $\;\pm\;$0.1684\;           
                                                             \\
                                                      & CD   & 0.6051 $\;\pm\;$0.1687\;    
                                                             & 0.6057 $\;\pm\;$0.1942\;        
                                                             & 0.5810 $\;\pm\;$0.1936\;         
                                                             & 0.3620 $\;\pm\;$0.1275\;          
                                                             & 0.3576 $\;\pm\;$0.1525\;     
                                                             & 0.3538 $\;\pm\;$0.1623\;           
                                                             \\
                                                (2018)& CS   & \textcolor{green}{0.5830 $\;\pm\;$0.1366\;}    
                                                             & 0.5895 $\;\pm\;$0.1312\;        
                                                             & \textcolor{blue}{0.6011 $\;\pm\;$0.1347\;}         
                                                             & 0.3141 $\;\pm\;$0.0850\;          
                                                             & \textcolor{green}{0.3292 $\;\pm\;$0.0937\;}     
                                                             & \textcolor{red}{0.3523 $\;\pm\;$0.1146\;}           
                                                             \\\midrule
			  \multirow{2}{*}{L-EMGNet}   & SD             & \textcolor{red}{0.9107 $\;\pm\;$0.0534\;}    
                                                             & \textcolor{green}{0.9254 $\;\pm\;$0.0550\;}        
                                                             & \textcolor{red}{0.9367 $\;\pm\;$0.0483\;}         
                                                             & \textcolor{red}{0.7051 $\;\pm\;$0.0783\;}          
                                                             & \textcolor{green}{0.7386 $\;\pm\;$0.0786\;}     
                                                             & \textcolor{red}{0.7917 $\;\pm\;$0.0740\;}           
                                                             \\
                                                      & CD   & \textcolor{red}{0.6803 $\;\pm\;$0.2099\;}    
                                                             & \textcolor{red}{0.7136 $\;\pm\;$0.1938\;}        
                                                             & \textcolor{red}{0.6905 $\;\pm\;$0.2065\;}         
                                                             & \textcolor{red}{0.4451 $\;\pm\;$0.1688\;}          
                                                             & \textcolor{red}{0.4548 $\;\pm\;$0.1855\;}     
                                                             & \textcolor{green}{0.4465 $\;\pm\;$0.1803\;}           
                                                             \\
                                                (2024 )& CS  & 0.5559 $\;\pm\;$0.1875\;    
                                                             & 0.5513 $\;\pm\;$0.1748\;        
                                                             & 0.5613 $\;\pm\;$0.1670\;         
                                                             & \textcolor{blue}{0.3143 $\;\pm\;$0.1151\;}          
                                                             & \textcolor{blue}{0.3290 $\;\pm\;$0.1039\;}     
                                                             & 0.3321 $\;\pm\;$0.1218\;           
                                                             \\
				\bottomrule

			\end{tabular}
			\begin{tablenotes}
				\item[*] The type SD, CD, and CS denote single-day, cross-day, and cross-subject experiments, respectively;
                    \item[*] The results highlighted in bold red/green/blue indicate that the model achieved the first/second/third ranking on the respective metric.
			\end{tablenotes}
		\end{threeparttable}
        }
\end{table*}

\section{Experiments and benchmark Results}
\subsection{Experiments setting}
To evaluate the performance of the gesture-free hand intention recognition system, we designed single-day, cross-day, and cross-subject experiments to test the existing sEMG classification algorithms. In the single-day experiment, the first eight blocks of data are used for training, while the rest are used for testing. On the one hand, it can avoid the class imbalance problem, and on the other hand, it ensures that the test data is collected after the training data to simulate actual application scenarios \cite{EMG_13}. The cross-day experiment uses the data from the first experiment as the training set to test the performance of the model in the second experiment. As for cross-subject experiments, Leave-One-Out cross-validation \cite{EMG_12} has been adopted to test the performance of the gesture-free hand intention recognition system. Considering that the requirements for interactive instructions vary in different application scenarios, the experiment is divided into 6 and 12 classes of finger force modes. Among them, the 6 classes of force modes only include resting state and single-finger force, while the 12 classes of force modes add 6 classes of combined force modes on the basis of the former. Through the literature review, the window sizes of sEMG signals were set to 250ms, 500ms, and 750ms, respectively, with a moving step size of 250ms to ensure real-time interaction response \cite{EMG_14}. All classification methods in our toolbox have been tested. Consistent with most studies, we only use accuracy to measure the performance of model classification.

\subsection{Benchmark Results}
Before analyzing the benchmark results, we used the Shapro-Wilk test \cite{EMG_15} to verify that the classification results on benchmark methods followed the normal distribution hypothesis. Therefore, we adopted mean ± standard deviation to record the experimental results. Tab. \ref{table_single_day} shows all classification results of single-day, cross-day, and cross-subjects with ten benchmark methods. 
\subsubsection{Results of single-day experiments}
focus on the first row of each classification method, the model L-EMGNet achieves the best performance with 250ms and 750 ms window size, and the 2DCNN achieves the best in 500ms. Their performance advantages relative to the second and third rank is as follows: 0.6\% and 2.35\% in the 6-classes experiment with a 250ms window. The gaps in 500ms and 750ms are 0.19\%, 2.55\% and 1.25\% and 3.17\%. In the 12-classes experiment, the advantages are 0.8\%, 0.53\%, and 3.75\% compared to the second rank and 4.12\%, 2.23\%, and 6.13\% compared to the third rank. Generally, the sEMG-based gesture-free hand intention recognition task achieves excellent classification performance, especially in 6-class experiments. 

\subsubsection{Results of cross-day experiments}
observe the second row, the model L-EMGNet achieves the best performance except for the 12-classes experiment with the 750ms window compared to Random Forests. The performance between the top-2 is relatively close, the gap is 1.18\%, 3.22\%, 0.44\%, 0.9\%, 0.29\%, and 2.32\%, respectively. The third rank of performance distributes scattered, including LDA, 2DCNN, and CNN-LSTM. The performance gap is 3.94\%, 5.64\%, 3.15\%, 4.49\%, 3.34\%, and 5.86\% compared to the best method. 

\subsubsection{Results of cross-subject experiments}
the third row of Tab. \ref{table_single_day} shows the cross-subject results. The results of L-EMGNet are no longer excellent, while the results of 2DCNN are better, achieving four first ranks and two second ranks. Specifically, the performance gap from the second rank to the best method is 2.13\%, 1.5\%, 0.54\%, 0.38\%, 0.35\%, and 0.06\%. Similar to the cross-day experiment, the third rank is distributed, appearing in four different methods. Compared to cross-day experiments, the performance gap is not significant, 2.36\%, 2.83\%, 1.81\%, 0.79\%, 0.37\%, and 0.38\%. 

\subsubsection{Comparative analysis}
comprehensively compare the results of single-day, cross-day, and cross-subject experiments, we have the following three observations:
\begin{itemize}
    \item [$\bullet$] The results of cross-day and cross-subject experiments have significantly decreased. Compared to single-day experiment, the performance gap based on the best method is 23.04\%, 21.37\%, 24.62\%, 26\%, 28.91\%, and 32.2\% in cross-day experiments and 30.63\%, 30.57\%,  31.75\%, 38.29\%, 41.12\% and 43.94\% in cross-subject experiments. This may be due to displacement deviation caused by two wearing attempts, as well as physiological differences among different subjects and on different days. 
    \item [$\bullet$] L-EMGNet achieved the best performance in the single-day and cross-day experiments but lagged behind 2DCNN in the cross-subject experiment. The reason may be that L-EMGNet is a lightweight model that performs well in experiments with small amounts of data. The cross-subject experiment involves a very large amount of data, which is more suitable for 2DCNN with a larger number of parameters. 
    \item [$\bullet$] Increasing the window size is beneficial for improving recognition accuracy, but it also increases the system's latency. Using the best methods as statistical objects, the performance of the 750ms window is higher than that of the 250 window is 2.6\%, 1.02\%, and  1.48\% in 6-classes experiments. The improvement is 8.66\%, 2.46\%, and 3.01\% in 12-classes experiments. This is because a longer time window is beneficial for obtaining more stable sEMG data.
    
\end{itemize}

\section{Discussion and Analyses}
To ensure fairness, all discussed algorithms use L-EMGNet, with data sourced from single-day experiments and sliced based on a 500ms window length.
\subsection{Evaluation of sEMG signal quality}
To evaluate the sEMG signal quality, we choose the following three metrics: (1) Signal-to-Noise Ratio (SNR), which is used to quantify the increase in sEMG signal amplitude during activation compared to background noise level recorded when the muscle is not contracting; (2) Signal-to-Motion Artifact Ratio (SMR): Motion artifacts are low-frequency noises that contaminate sEMG signals and are introduced by fine electrode movements and changes in contact at the electrode–skin interface; and (3) Power Spectrum Deformation ($\Omega$): which is sensitive to symmetry and peaking in the power spectrum and to additive disturbances in the low- and high-frequency ranges. The calculation formula for all metrics is as follows:
\begin{equation}
    \begin{split}
        \text{SNR} = 20\text{log}(\text{$\text{rms}_{\text{Activation$\_$filtered}}$/$\text{rms}_{\text{Resting$\_$filtered}}$}) \\
        \text{SMR} = 10\text{log}([\sum\limits_{f = 0}^{500} {{\rm{PS}}{{\rm{D}}_{{\rm{filtered}}}}/\sum\limits_{f = 0}^{20} {{\rm{PS}}{{\rm{D}}_{{\rm{AboveLine\_raw}}}}} }]) \\
        \Omega  = 10\log [\frac{{{{({{\rm{M}}_2}/{{\rm{M}}_0})}^{1/2}}}}{{{{\rm{M}}_1}/{{\rm{M}}_0}}}], \;
        {{\rm{M}}_i} = \sum\limits_{f = 0}^{500} {{\rm{PSD }}{\rm{.}}{f^i}} \qquad\\
    \end{split}
\end{equation}

Regarding force mode 1 as a resting state, we calculated the SNR, SMR, and PSD of five other single-finger force modes, the results shown in Tab. \ref{tab:quality}. The results in Botros et al. \cite{EMG_8} research are also included in the table as a reference to determine the quality of our collected sEMG data. On the most critical SNR metric, the results are relatively close, but on SMR and  $\Omega$ metric, the results are relatively poor, indicating that the data contains severe motion artifacts, possibly due to our data being collected in multiple motion modes. In addition, we found that the signal quality of the middle finger, ring finger, and little finger is better, which may be due to the fact that when holding a cylindrical object, its gravity will involuntarily drive the thumb and index finger to exert force. 

\begin{table}[htb]
    \centering
    \caption{Signal Quality of the Collected sEMG Data, With the 6-classes Force Modes. }
    \resizebox{0.5\textwidth}{!}{
        \begin{threeparttable}
            \begin{tabular}{c|ccc}
            \toprule
             \textbf{Mode}  & \textbf{SNR}$\;\uparrow$      & \textbf{SMR} $\;\uparrow$   &\textbf{$\Omega$} $\;\downarrow$ \\
             \midrule
             \textbf{mode 2} & 12.3 (6.6)\;/\;8.0  (3.6)    & 11.9 (6.2)\;/\;1.5 (2.8)   & 1.04 (0.64)\;/\;3.10 (0.52) \\
             \textbf{mode 3} & 11.5 (4.7)\;/\;8.3  (2.4)    & 11.4 (5.2)\;/\;1.6 (2.3)   & 1.03 (0.48)\;/\;3.00 (0.41) \\
             \textbf{mode 4} & 7.8  (4.9)\;/\;\textbf{12.6 (3.7)}    & 7.8  (5.2)\;/\;5.0 (2.7)   & 1.34 (0.89)\;/\;2.24 (0.54) \\
             \textbf{mode 5} & 11.4 (5.2)\;/\;\textbf{14.0 (3.4)}    & 11.2 (4.8)\;/\;6.4 (3.4)   & 0.97 (0.42)\;/\;2.05 (0.60) \\
             \textbf{mode 6} & 11.0 (4.2)\;/\;\textbf{11.8 (3.9)}    & 10.7 (4.3)\;/\;4.2 (3.4)   & 0.98 (0.36)\;/\;2.48 (0.65) \\
             \bottomrule
            \end{tabular}
            \begin{tablenotes}
                \item[*] The results on the left are from \cite{EMG_8}, and the results on the right are from this study;
                \item[*] Higher values of SNR and SMR and lower values of $\Omega$ are better.
            \end{tablenotes}
        \end{threeparttable}
    }
    \label{tab:quality}
\end{table}

\subsection{Details of the classification results}
\begin{figure}[htb]
    \centering
    \subfigure[6-classes]{\includegraphics[width=0.49\columnwidth]{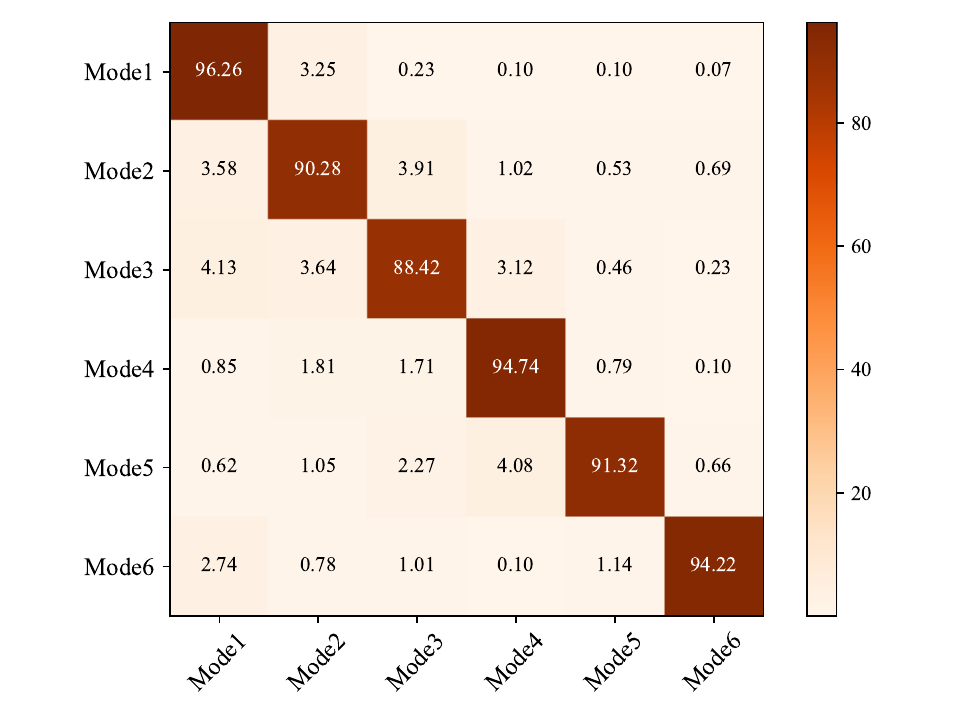}}
    \subfigure[12-classes]{\includegraphics[width=0.49\columnwidth]{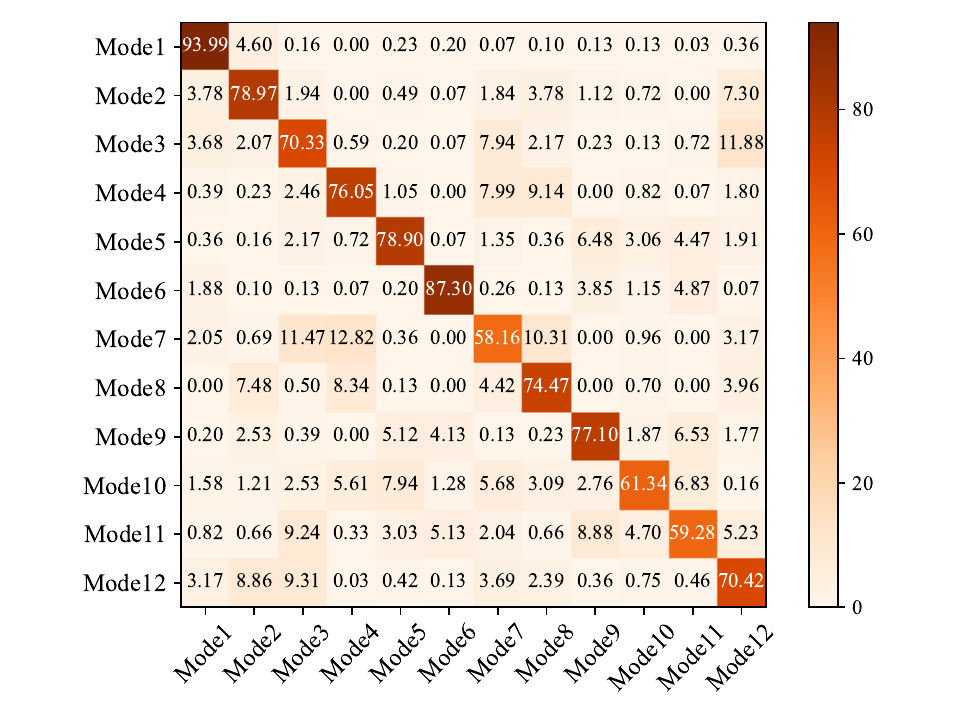}}
    \caption{Confusion matrix in single-day experiments with the 500ms window size using L-EMGNet. (a) 6-classes force modes; (b) 12-classes force modes.}
    \label{fig-cm}
\end{figure}

It is important to explore the classification details of the gesture-free hand intention recognition task, this is beneficial for screening out poor force modes, thereby improving the robustness of the system. The confusion matrices for 6 and 12 classes on a single-day experiment, as shown in Fig. \ref{fig-cm}. Observations have shown that the accuracy of patterns that include the use of index finger force is relatively low. For example, in the single-finger experiment of class 6, the index finger performed the worst. In the combined finger experiment, the results of the third, seventh, eleventh, and twelfth force modes were poor, while patterns that do not include index finger force were relatively good, except for the eighth force mode. This suggests that in future research, we can increase the use of the middle finger, ring finger, and little finger, and reduce the use of the index finger. 

\subsection{The factor of different speeds}
Considering multiple application scenarios, it is necessary to explore the interaction accuracy of the system in different motion modes. To achieve this purpose, we recalculate the results according to its sports speeds, e.g. 0, 4, 6, 8 km/h. As illustrate in Tab. \ref{tab:speed}, we found that there is no significant difference in classification performance among various motion modes. Two reasons are considered: on the one hand, deep learning has the ability to extract features from sEMG signals containing motion noise; On the other hand, motion noise is mainly distributed in low frequencies (below 20Hz) and is filtered out in the preprocessing stage.

\begin{table}[htb]
    \centering
    \caption{Performance with Different Sports Speed. }
    \begin{threeparttable}
        \begin{tabular}{c|cc}
        \toprule
         \textbf{Speed}  & \textbf{6-classes}   & \textbf{12-classes} \\
         \midrule
         \textbf{0 km/h} & 0.9279 $\;\pm\;$ 0.0521       &  0.7360 $\;\pm\;$ 0.0879 \\
         \textbf{4 km/h} & 0.9251 $\;\pm\;$ 0.0583       &  0.7386 $\;\pm\;$ 0.0851 \\
         \textbf{6 km/h} & 0.9238 $\;\pm\;$ 0.0597       &  0.7383 $\;\pm\;$ 0.0696 \\
         \textbf{8 km/h} & 0.9248 $\;\pm\;$ 0.0567       &  0.7416 $\;\pm\;$ 0.0797 \\
         \textbf{Mixed}  & 0.9254 $\;\pm\;$ 0.0550       &  0.7386 $\;\pm\;$ 0.0786 \\
         \bottomrule
        \end{tabular}
    \end{threeparttable}
    \label{tab:speed}
\end{table}
\subsection{The factor of different force intensities}
\begin{figure}[htb]
    \centering
    \centerline{\includegraphics[width=0.8\columnwidth]{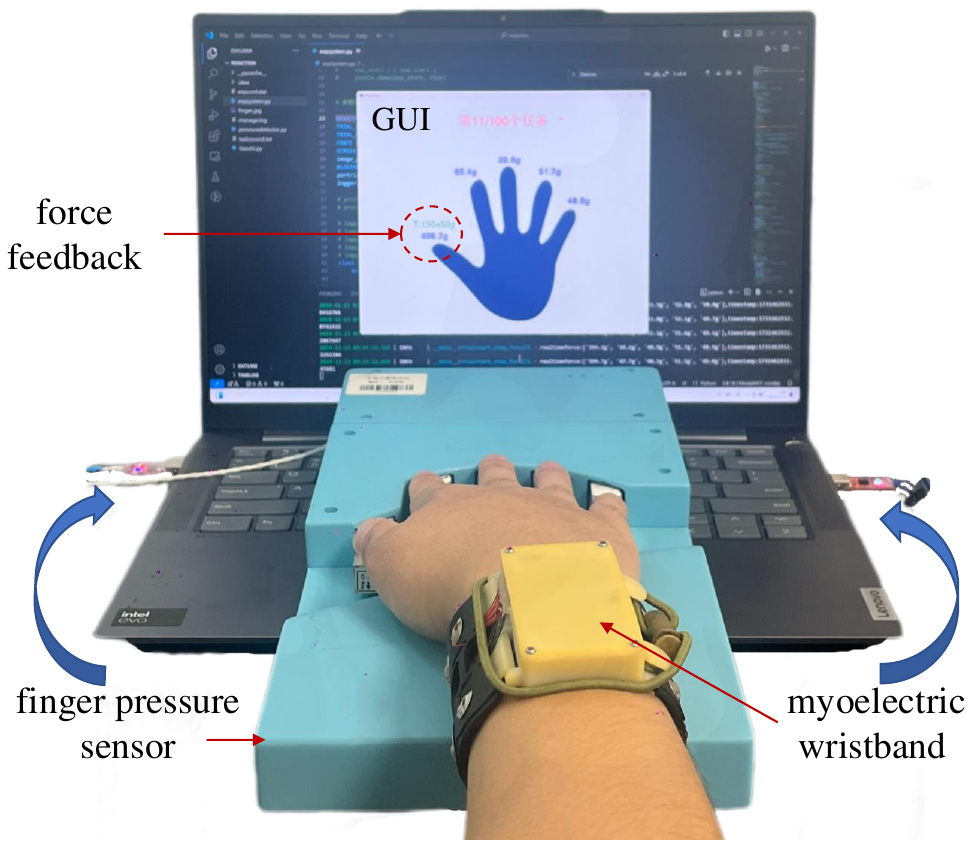}}
    \caption{Schematic diagram of the force pressure test system. }
    \label{fig-pressure}
\end{figure}

\begin{figure}[htb]
    \centering
    \centerline{\includegraphics[width=0.8\columnwidth]{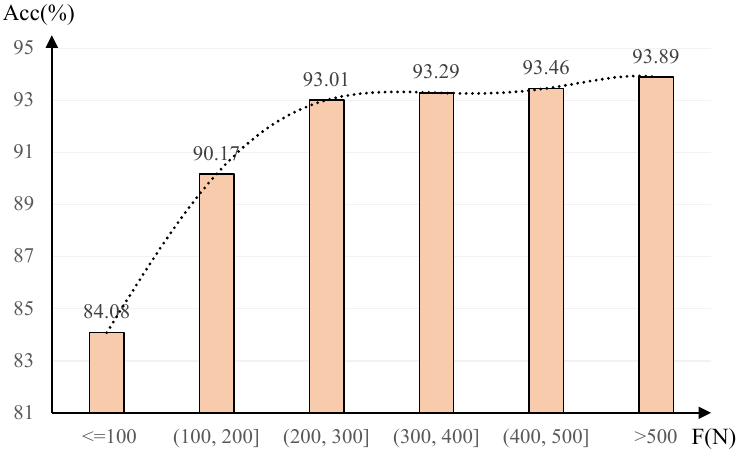}}
    \caption{Average results of each level force intensities. The curve is fitted by a 5th-order polynomial.}
    \label{fig-result_p}
\end{figure}

Different finger force intensities may affect the recognition accuracy of the system, which is worth exploring for their relationship. As shown in Fig. \ref{fig-pressure}, we have designed a force sensor to measure the force intensity of each finger. Considering that it is difficult to control force, the experiment only includes 6-classes of experiments using a single finger. Experiments were conducted on three subjects, S04, S06, and S08, which are the best, average, and worst subjects in benchmark experiments. The force intervals include as follows: [0, 100N], (100N, 200N], (200N, 300N], (300N, 400N], (400N, 500N], (500N, +$\infty$). The accuracy of system recognition will be calculated based on the interval to which the force belongs. As shown in Fig. \ref{fig-result_p}, the accuracy continues to improve and the trend of improvement gradually slows down with the finger force increases. This is because after exerting a certain degree of force, muscles have already contracted in place.

\subsection{Online experiments}
\begin{figure}[htb]
    \centering
    \centerline{\includegraphics[width=0.9\columnwidth]{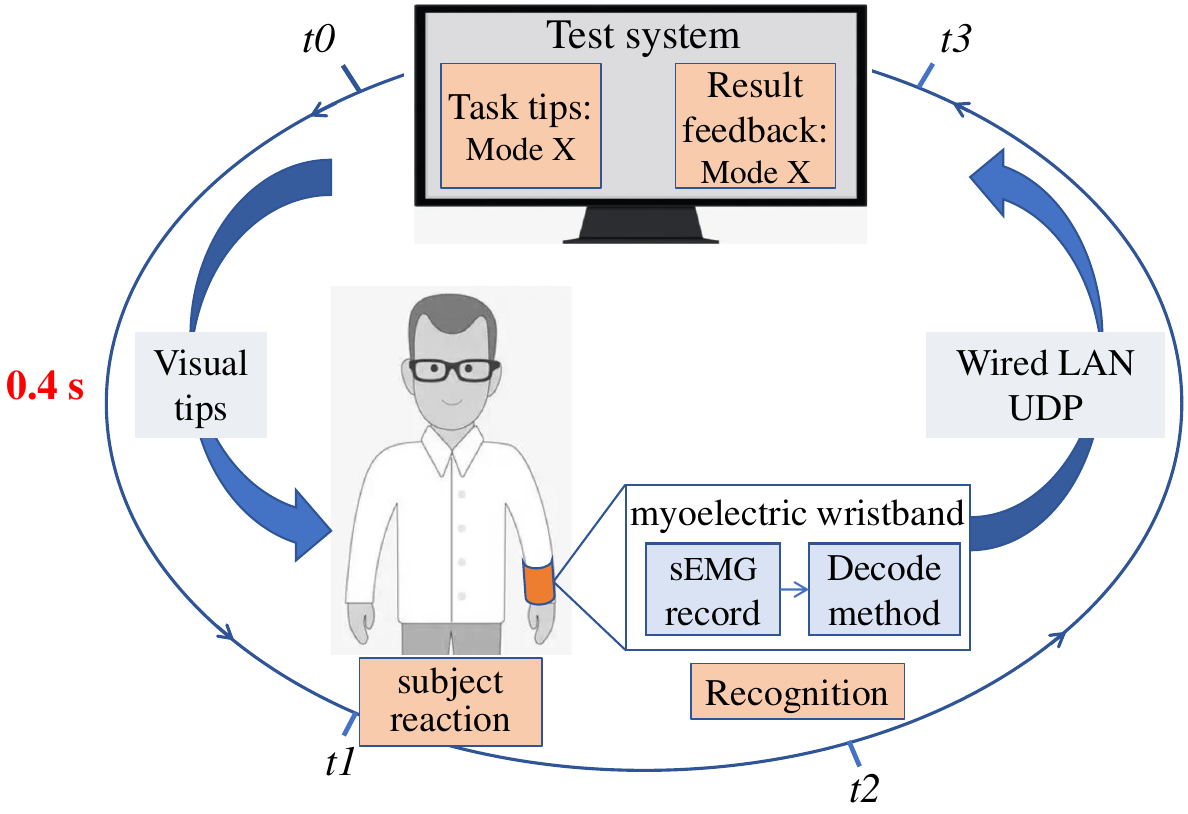}}
    \caption{Schematic diagram of the online experiment system. }
    \label{fig-online}
\end{figure}

Accuracy and system response time are the two most important aspects of online experiments. This experiment is based on the same three subjects, and the testing system is shown in Fig. \ref{fig-online}. Based on the keyboard experiment, the average reaction time of the three subjects was about 0.4s, which denotes the time from visual tips to the subject's reactions and should not be included in the system's reaction time. Thus, the system response time can be euqated as $\Delta t = {t_3} - {t_1} \approx {t_3} - {t_0} - \text{0.4s}$, because the $t_1$ is unable to accurately calculate. Each subject must complete 50 random force mode tests. Each test randomly prompts one force mode except for the resting state, and the first non-resting state prediction of the system is the predicted result and completion time, i.e. $t3$. The accuracies of the three subjects are 94\%, 90\%, and 88\%, and system response times are 0.35s, 0.31s, and 0.3s, respectively. The excellent online results further validated the robustness in practical applications. 
\section{Conclution}
This work proposes the sEMG-based gesture-free hand intention recognition, which is a novel scheme for covert message transmission between collaborators in scenarios where collaborators and non-collaborators coexist. This scheme hides the action of sending messages, making it difficult for non-collaborators to detect, thereby protecting privacy and information security. To verify the feasibility, we designed the experimental system and paradigm, collected the dataset, and developed the signal processing toolbox, ultimately obtaining benchmark results on various decoding methods. All data, software, and hardware designs and code had been open-sourced to promote the development of this research field. The benchmark results and discussion analysis indicate that the system performs well in a single-day experiment, but needs to be improved in cross-day and cross-subject experiments. Therefore, methods such as domain adaptation and testing adaptation that are suitable for solving distribution shift problems will be attempted in the future. 

\newpage
\bibliographystyle{IEEEtran}
\bibliography{IEEEabrv, appendix}

\end{document}